# Clinical Validation of Single-Chamber Model-Based Algorithms Used to Estimate Respiratory Compliance

Gregory B. Rehm MS*
Department of Computer Science, University of California Davis
1 Shields Ave.
Davis CA, 95616
grehm@ucdavis.edu

Jimmy Nguyen*
Department of Respiratory Care, University of California Davis Health System
jinguyen@ucdavis.edu

Chelsea Gilbeau
Department of Respiratory Care, University of California Davis Health System
cmgilbeau@ucdavis.edu

Marc T Bomactao
Department of Respiratory Care, University of California Davis Health System
mbomactao@ucdavis.edu

Chen-Nee Chuah, PhD
Department of Computer and Electrical Engineering, University of California Davis
1 Shields Ave.
Davis CA, 95616
chuah@ucdavis.edu
530-752-5825

Jason Adams, MD MS
Division of Pulmonary and Critical Care Medicine, University of California Davis
4150 V Street, Suite 3400
Sacramento, CA 95817
jyadams@ucdavis.edu
916-734-3565

**Corresponding author:** Jason Y. Adams, MD MS

*authors contributed equally to this work

# ABSTRACT

Non-invasive estimation of respiratory physiology using computational algorithms promises to be a valuable technique for future clinicians to detect detrimental changes in patient pathophysiology. However, few clinical algorithms used to non-invasively analyze lung physiology have undergone rigorous validation in a clinical setting, and are often validated either using mechanical devices, or with small clinical validation datasets using 2-8 patients. This work aims to improve this situation by first, establishing an open, and clinically validated dataset comprising data from both mechanical lungs and nearly 40,000 breaths from 18 intubated patients. Next, we use this data to evaluate 15 different algorithms that use the "single chamber" model of estimating respiratory compliance. We evaluate these algorithms under varying clinical scenarios patients typically experience during hospitalization. In particular, we explore algorithm performance under four different types of patient ventilator asynchrony. We also analyze algorithms under varying ventilation modes to benchmark algorithm performance and to determine if ventilation mode has any impact on the algorithm. Our approach yields several advances by 1) showing which specific algorithms work best clinically under varying mode and asynchrony scenarios, 2) developing a simple mathematical method to reduce variance in algorithmic results, and 3) presenting additional insights about single-chamber model algorithms. We hope that our paper, approach, dataset, and software framework can thus be used by future researchers to improve their work and allow future integration of "single chamber" algorithms into clinical practice.

**Funding Sources:** This study was funded by NIH 5F31HL144028-03. The authors do not have conflicts of interests to disclose. The content is solely the responsibility of the authors and does not necessarily represent the official views of the NIH.

# INTRODUCTION

Patients who are receiving mechanical ventilation in the intensive care unit (ICU) are heavily monitored for changes in breathing and respiratory function[1], [2]. One such lung physiologic parameter that is monitored is respiratory compliance ($C_{rs}$). $C_{rs}$ measures the ease at which the lung is able to stretch when filling with air[3], and its measurement allows optimization of ventilator settings, warning of deterioration in respiratory pathophysiology, and possibly even warning of onset of lung diseases such as the Acute Respiratory Distress Syndrome (ARDS)[4], [5]. Unfortunately, monitoring of $C_{rs}$ is not easily performed in the ICU because it involves time consuming procedures[6]–[9], or use of clinically invasive devices that may adversely affect care[10]. These barriers mean that $C_{rs}$ can go unrecorded and thus unused when making important care decisions for a patient, which in turn may affect patient care outcomes[5], [11]–[13].

To offer solution to this problem, a non-invasive and continuous measurement of respiratory mechanics has been proposed by modeling the lung as a single compartment governed by the equation of motion[6], [14]. This approach makes simplifying assumptions about the behavior of the lung that has allowed researchers to measure $C_{rs}$ on mechanically ventilated patients using a number of different computer algorithms[5], [7], [14]–[24]. Many of these algorithms have been validated in idealized circumstances, however, few of them have been thoroughly tested in scenarios typically seen in the clinical environment[10], [15], [25].

The clinical environment often poses a number of challenges to $C_{rs}$ estimator algorithms because there are various type of breathing and ways that a breath can be delivered to a patient that cause estimation error. One scenario in which algorithms may provide inaccurate or inconsistent estimates is- in cases of severe PVA (Figure 1B). PVA occurs when patient breathing is mismatched with ventilator settings[2], which can result in breathing patterns that $C_{rs}$ estimators are not designed to handle. Another scenario that $C_{rs}$ estimators may struggle to adapt to is different ventilation modes (VMs)[7]. VM determines the manner in which a patient receives a breath from the ventilator. One mode, named volume control (VC) specifies that patients receive a fixed amount of air per breath. Other modes such as pressure control (PC) and pressure regulated volume control (PRVC) are designed so that patients receive a fixed amount of pressure during a breath and can breathe a more flexible amount of air per breath. A number of algorithms have been developed specifically for VC[21], [22], while others have been tested and validated under PC or other VMs[7], [15], [25]. However, it is unclear if these algorithms are cross-compatible with other VMs because the research community lacks widely available data sources from a variety of patients, and because of the difficulty in collecting ventilation data during clinical trials. This lack of validation also makes it unclear to clinicians and researchers which algorithm can be used in varying circumstances, and may hinder the implementation of model-based approaches in clinical practice.

To address this situation, we explore the performance of a comprehensive number of algorithms for estimating $C_{rs}$ in a retrospective cohort of subjects under varying clinical scenarios. First, we explore general algorithmic performance across all patients, and then we focus on performance under scenarios of differing VM and types patient-ventilator synchrony (PVA). In doing so, we highlight several insights. First, we highlight which algorithms are best for calculating $C_{rs}$ under varying scenarios of patient state such as varying VMs and PVAs. Second, we discuss a windowing approach that will help improve usage of algorthmic results in clinical practice. Third, we show a surprising result that algorithms designed for VC perform equally well when applied to patient breath data collected from mechanical ventilators operating under pressure modes, and vice versa. These contributions and our open dataset will help improve the understanding of advantages and limitations of the model-based approach and may help guide clinicians and researchers in using $C_{rs}$ estimators in their own practice and studies.

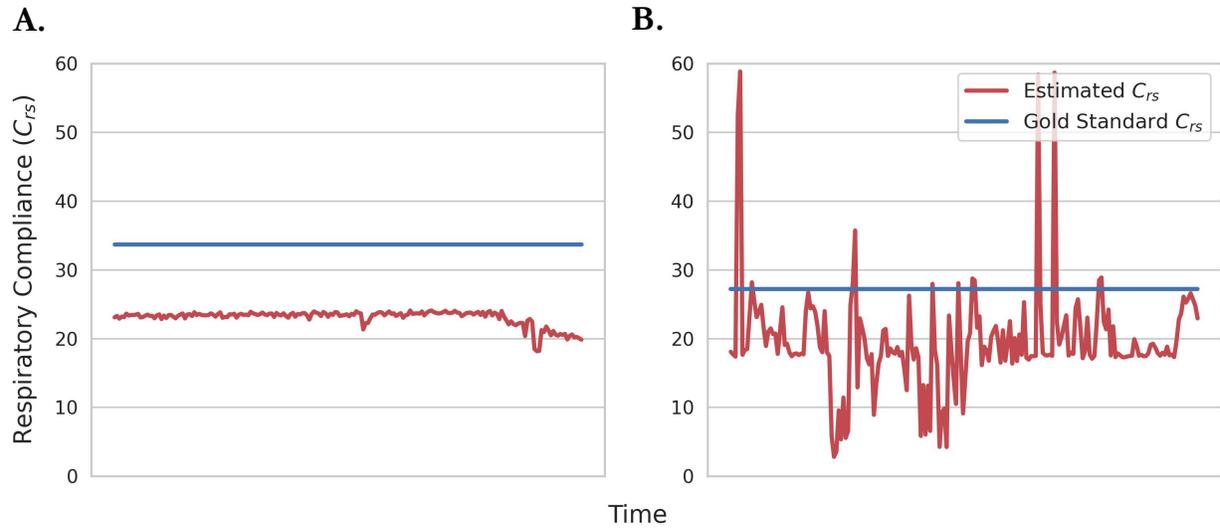

Figure 1: We provide 2 samples of data from different patients. In **A.** the patient is breathing synchronously with the ventilator and there is little variation in $C_{rs}$ estimates. In **B.** the $C_{rs}$ estimate is highly variable due to patient ventilator asynchrony or transient anomaly.

# MATERIAL AND METHOS

## Subject Cohort Selection

This is a retrospective study that was conducted as part of an institutional review board (IRB)-approved research effort collecting ventilator waveform data (VWD) from mechanically ventilated adults at UC Davis Medical Center (UCDMC). Enrolled subjects were considered for analysis if a clinically valid inspiratory pause maneuver[6] was performed. All subjects were ventilated in the following assist control VMs: VC, PC, or PRVC. Subjects had RASS scores ranging from -1 to -5. Subjects were excluded from analysis if they were under 18 years of age, were pregnant or a prisoner, or if all plateau pressure checks were determined invalid.

## Data Collection

We used a previously published methodology to collect VWD[26]. VWD consists of raw pressure (cm $H_2O$) and airflow (L/min) observations sampled at 50 Hz. Our dataset is split into two parts: first we collected VWD using a mechanical lung (QuickLung$^{TM}$) over the course of 8 different experiments, and 8,787 breaths. We used this data to validate all reproduced algorithm code. Next, we collected a dataset of 39,660 breaths from 18 subjects hospitalized at UCDMC between 2015 and 2020. Ground-truth $C_{rs}$ was found for each breath through analysis of end-inspiratory pause maneuvers that were performed during routine course of care[5], [27]. All inspiratory pause maneuvers were validated by three clinicians (JN, CG, MB) by either performing the procedure at bedside, or through retrospective review of VWD. VMs and RASS scores were collected via review of both electronic medical record (EMR) and collected VWD[28]. All VWD was processed through rule-based algorithms to determine presence of the following subtypes of PVA: double trigger asynchrony, breath stacking asynchrony, flow asynchrony (FA), and delayed cycling asynchrony (DCA)[29]. Flow asynchrony is further graded by severity in terms of the level of inspiratory pressure deflection into mild, moderate and severe grades of asynchrony.

**Algorithms Evaluated**

We replicated 15 different computer algorithms used for non-invasively estimating $C_{rs}$ using the single-chamber model of the lung[30], [31]. The single-chamber model is expressed by:

$$P_{aw}(t) = R_{aw}Q(t) + \frac{V(t)}{C_{rs}} + P_0,$$

where $P_{aw}(t)$ is the airway pressure at a given time t, $R_{aw}$ is the patient's airway resistance, $Q(t)$ is airflow at time t, $V(t)$ is the volume of air in the lung at time t, $C_{rs}$ is the respiratory compliance, and $P_0$ is the positive end expiratory pressure (PEEP). Although analytic methods exist[15], [17], this equation is usually solved by using least squares regression to calculate for $C_{rs}$ and $R_{aw}$ terms using measured $Q(t)$ and $V(t)$ data from either the inspiratory or expiratory limbs of a breath[7], [14], [23]. This method is able to non-invasively measure lung physiology and can accurately estimate $C_{rs}$ when patients are exerting no muscular effort in breathing.

The single chamber model, however, fails to account for muscular effort during spontaneous breathing. Furthermore, the performance of $C_{rs}$ estimators during spontaneous breathing has received scant validation[10]. So, in this paper, we analyze a comprehensive list of $C_{rs}$ estimation methods including: Al-Rawas method [17], [32], flow-targeted inspiratory least squares[7], Howe's least-squares[23], IIPR[21], IIMIPR[33], IIPREDATOR[10], Kannangara's method[7], Major's method[24], MIPR[33], polynomial method[34], PREDATOR[22], pressure-targeted expiratory least-squares[5], pressure-targeted inspiratory least squares[30], [31], constrained optimization[16], and Vicario's non-invasive estimation of alveolar pressure[15]. Many of these methods are mode-restricted where they are designed to operate in either VC or PC/PRVC modes (Table 1). So, in our VM experiments, we only evaluated algorithms that were designed to work for a specific mode. For instance, in VC mode experiments, we only evaluated VC-specific algorithms, and mode-agnostic methods. Other algorithms, such as Howe's least squares[23] should theoretically work in any VM, but are unvalidated in VC.

| VC Algorithms | PC/PRVC Algorithms | Mode-Agnostic Algorithms | Mode-Agnostic Algorithms but Unvalidated in VC |
|---|---|---|---|
| IIPR[21] | Flow-targeted inspiratory least squares[7] (Figure 2C) | Al-Rawas method[17], [32] | Howe's least-squares[23] |
| IIMIPR[33] | Kannangara's method[7] | Pressure-targeted expiratory least-squares[5], [23] (Figure 2B) | Vicario's non-invasive estimation of alveolar pressure[15] |
| IIPREDATOR[10] | | Constrained optimization[10], [16] | |
| Major's method[24] | | | |
| MIPR[33] | | | |
| Polynomial method[34] | | | |
| PREDATOR[22] | | | |
| Pressure-targeted inspiratory least squares[5] (Figure 2A) | | | |

Table 1: Enumerates list of mode-specific, and mode-agnostic algorithms we analyze in this work.

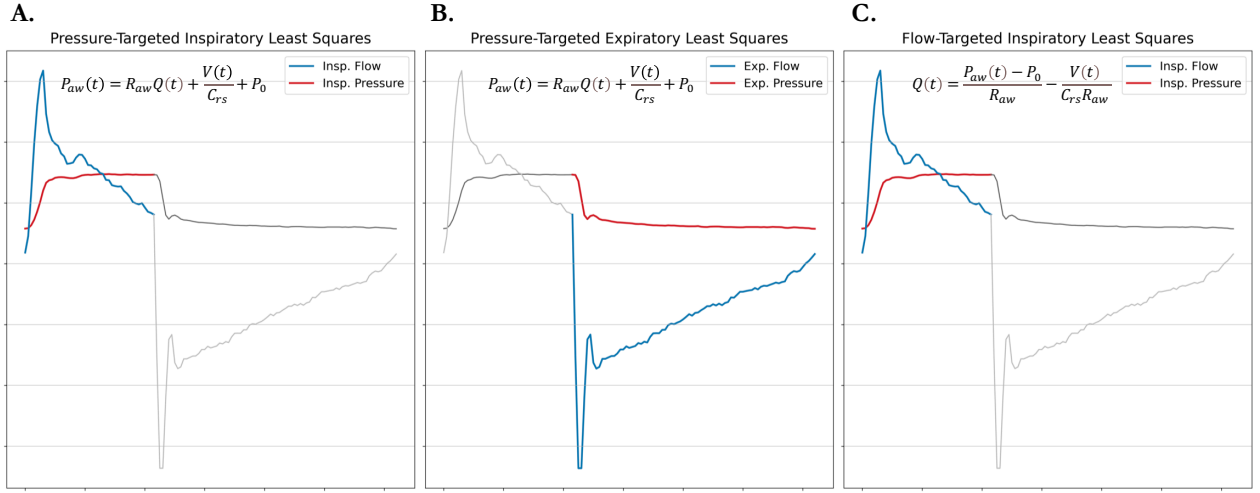

Figure 2: Shows different least squares regression methods and what kinds of data they use from different parts of the breath. Also shows which formulation of the single-chamber model is used in calculation.

## Metrics

We define respiratory compliance $C_{rs}^k$ as the measured compliance of breath $k$, $1 \geq k \geq N$ given $N$ breaths in any given patient sample. We measure $C_{rs}^k$ by performing an inspiratory pause maneuver to find plateau pressure, and then we derive $C_{rs}^k$ from the plateau pressure ($P_{plat}^k$) [6] using the equation:

$$C_{rs}^k = \frac{V^k}{P_{plat}^k - P_0}$$

where $V^k$ is the inhaled volume of air for breath $k$. We make the assumption that $C_{rs}$ does not change significantly in a short period of time [10], so we assume that all breaths within 30 minutes of the time of finding $P_{plat}^k$ have equivalent $C_{rs}^k$. If more than one plateau pressure was taken within 30 minutes of a breath, then breath $C_{rs}^k$ was averaged across the multiple observations.

After our gold-standard compliance $C_{rs}^k$ for each breath is found, we can compute the absolute difference ($AD^k$), and standard difference ($D^k$), between $C_{rs}^k$ and the estimated compliance $\hat{C}_{rs}^k$ for each breath from a chosen algorithm.[10], [23], [24]

$$AD^k = |C_{rs}^k - \hat{C}_{rs}^k|$$

We can then find various statistics, such as median, mean, and median absolute deviation (MAD) [10], [23], for all $AD^k$, $1 \geq k \geq N$ for any given patient. This allows us to determine performance of any algorithm for a given patient. Per-breath AD can be highly variable however, due to periods of PVA and transient abnormality, and may not always be clinically usable by clinicians. To ameliorate this issue, we propose to measure algorithmic performance by using windowing for a given window size $s$, and then smooth $\hat{C}_{rs}^k$ using the median of $\hat{C}_{rs}^k$ and $s - 1$ prior samples. We define this metric as windowed difference (WD):

$$WD^k = |C_{rs}^k - median(\{\hat{C}_{rs}^{k-s}, \dots, \hat{C}_{rs}^k\})|$$

WD can be calculated using a series of overlapping, or non-overlapping windows. For this paper we use overlapping windows with $s = 100$.

**Reporting**

We use the metrics defined above to examine how the algorithms perform across all patients in our dataset. For this we compute the median AD for each patient in our dataset, per algorithm. Then we compute the mean of each patient median across all patients to generate a mean AD across all patients. Mean MAD is computed in a similar way. The MAD of estimated compliance is found per patient, and then the mean MAD is computed across all patients per algorithm. These two computations are motivated by our desire to have an algorithm that on average estimates patient $C_{rs}$ as accurately as possible, while ensuring that $C_{rs}$ estimates are consistent.

There are also algorithm statistics that are best reported on a per-breath basis. For per-breath performance we collect all standard differences $D^k$ between estimated and true $C_{rs}$ across all patients and then plot results using a bar and whiskers plot. This allows us to examine algorithm performance in specific circumstances, such as during asynchronous breathing, when utilizing per-patient results not representative of patient-level clinical realities.

# RESULTS

## Algorithm Performance by Mode and Asynchrony Types

First, we compute the mean MAD of the best performing algorithms for VC mode and PC/PRVC modes (Figure 3, Figure 4) on a per-patient basis. We then perform a similar analysis across all VMs using only mode-agnostic algorithms. Our results show that most mode-agnostic algorithms underperform mode-specific algorithms. The best mode agnostic algorithm is Vicario's non-invasive estimation of alveolar pressure. We find however that this algorithm is highly variable and has the largest mean MAD compared to all other algorithms surveyed.

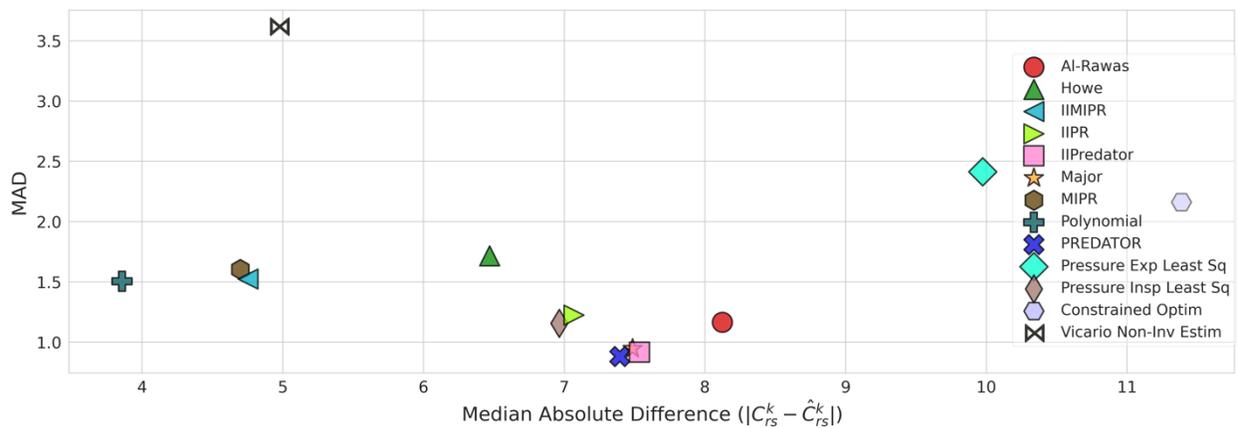

Figure 3: Displays which algorithm performs best on a per-patient basis for all volume control (VC) breaths. X-axis is found by computing the mean, median absolute difference between gold standard compliance and algorithm estimated compliance on a per-patient basis. Y-axis shows the mean median absolute deviation of all compliance estimates per patient. Note that VC specific algorithms and mode-agnostic algorithms are included in this analysis.

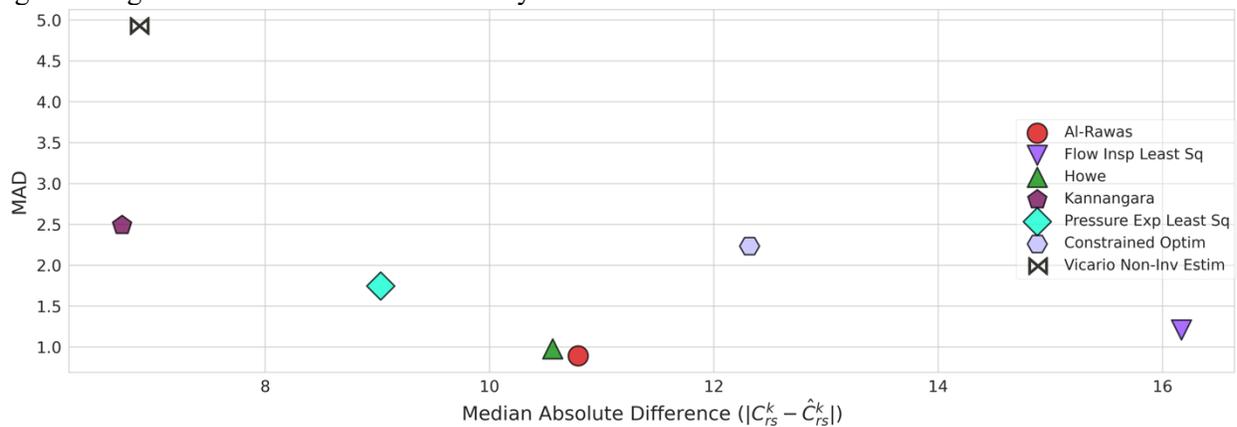

Figure 4: Displays which algorithm performs best on a per-patient basis for all pressure control (PC) and pressure regulated volume control (PRVC) breaths. Note that PC/PRVC specific algorithms and mode-agnostic algorithms are included in this analysis.

We also measure algorithmic performance under different PVA scenarios across VMs *using per-breath analysis*. Here we use box charts to describe the distribution, median, and IQR of standard differences between estimated and true $C_{rs}$. We do not use absolute difference for boxplots because standard difference allows us to better see distribution of errors around 0. Figure 5 shows the performance of VC-capable algorithms during asynchronous breathing excluding mild FA. We exclude mild FA because we found almost all algorithms are not negatively affected by it, and the preponderance of mild-FA in our dataset dampens the effect of other PVA types on our results. Our results in Figure 5 further highlight our findings from Figure 3 that VC VWD is best analyzed by VC specific algorithms, and that mode generic algorithms, with exception of Al-Rawas method, are less performant on VC breathing. Al-Rawas itself has a mechanism for filtering breathing that does not have a ideal exhalation curve, which be part of the reason why Al-Rawas performs well with asynchronous breathing in VC.

In Figure 6, we investigate the effect of moderate to severe FA on VC performance. This investigation is based on the hypothesis that VC algorithms work best in this scenario because they can correct for inherent nature of FA. We find that this hypothesis is partially correct, and that methods used to correct for FA, like IIPREDATOR, perform well in FA. Our analysis also finds that algorithms using expiratory data, instead of inspiratory data, like Howe's expiratory least squares are actually best at estimating $C_{rs}$ during moderate-severe FA. This may be because the expiratory limb is less affected than the inspiratory limb during FA.

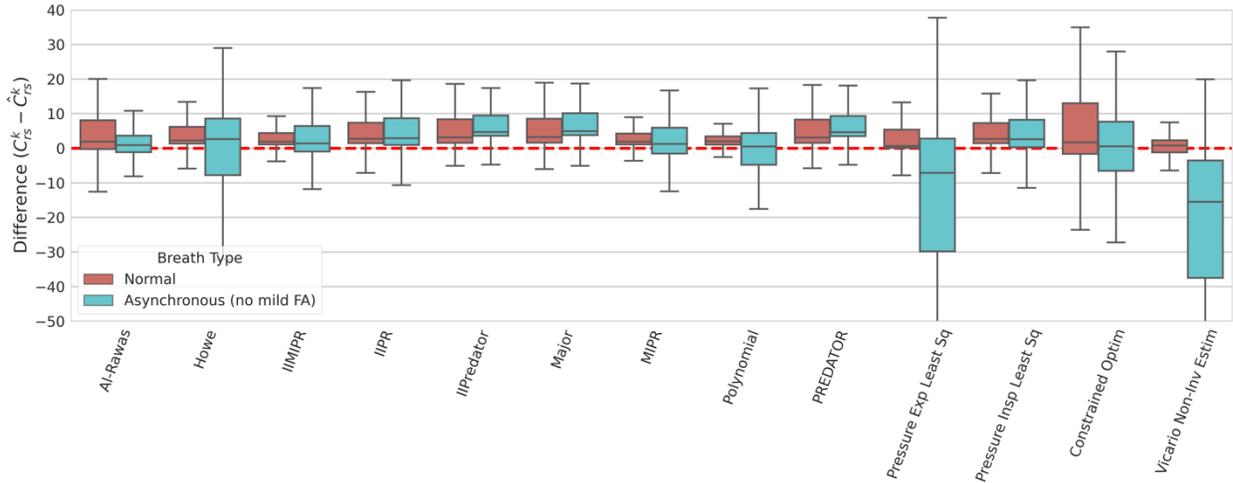

Figure 5: Shows boxplot comparison between algorithm performance between non-asynchronous and asynchronous breathing (without mild flow asynchrony) in volume control (VC) mode.

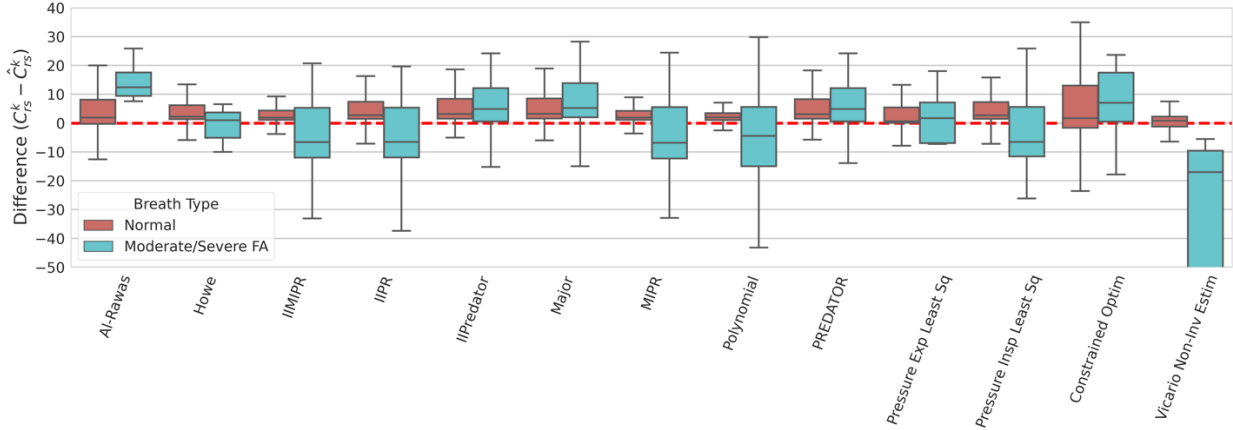

Figure 6: Shows boxplot comparison of algorithm performance between non-asynchronous and moderate/severe flow asynchrony (FA) breathing in volume control mode. We cut off results of Vicario's non-invasive estimation of alveolar pressure because the IQR would disrupt visualization of results.

In Figure 7 we show investigation of the prior methodology comparing normal and asynchronous breathing, but applied to PC/PRVC only. We see that almost all algorithms are relatively unaffected by PVA. However, like VC, this is because milder effects from DCA mutes effects of more severe asynchronies. When we remove DCA we find all algorithms perform significantly worse at estimating $C_{rs}$ than during normal breathing (Figure 8). We also find the Kannangara method tends to yield median differences closest to 0 of all algorithms surveyed. This is not surprising because the Kannangara had the best per-patient results for PC/PRVC breaths (Figure 4) and it was designed to correct for PVA.

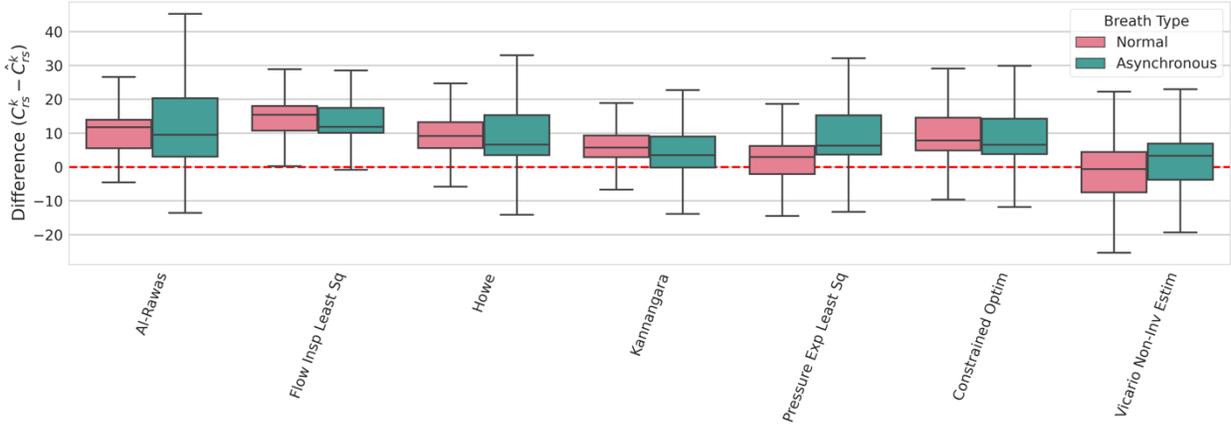

Figure 7: Boxplot comparison of algorithm performance between normal and asynchronous breathing in pressure modes (pressure control / pressure-regulated volume control).

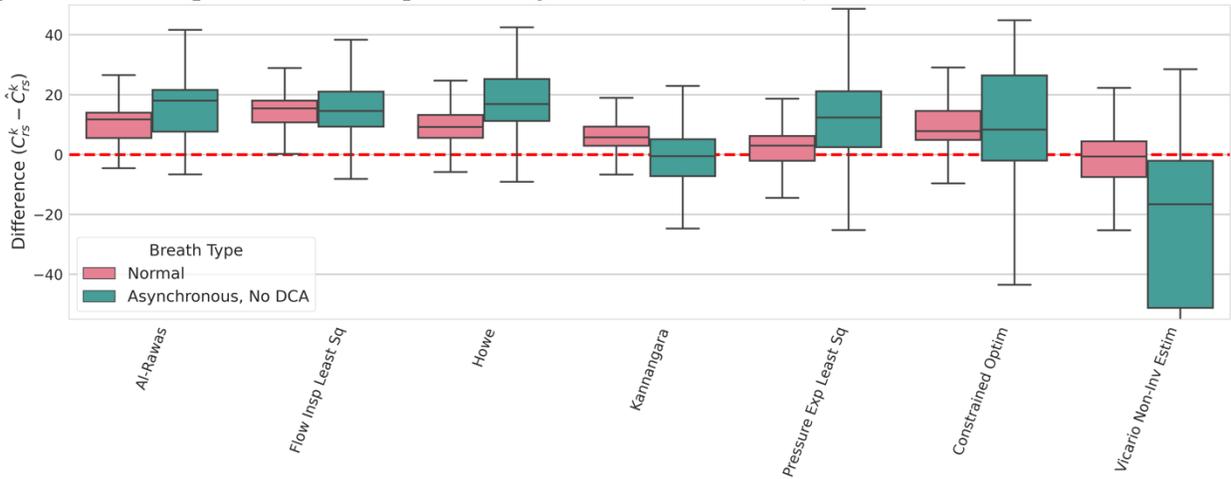

Figure 8: Boxplot comparison of algorithm performance between normal and asynchronous breathing without delayed cycling asynchrony (DCA) in pressure modes.

**Using Windowing**

We note that prior results (Figure 5-Figure 8) show per-breath IQR increases significantly during PVA. This finding has motivated development of algorithms that are resistant to PVA[7], [21], [22], [34], and has motivated our hypothesis that windowed difference would function to improve algorithm resistance to PVA and other transient anomalies in breathing. To investigate, we first analyze results of windowing on a per-patient basis. We found that for all algorithms, windowing did not yield any significant change in AD (Figure 9A), however, windowing does improve per-patient algorithm MAD for a majority of surveyed algorithms (Figure 9B). We also investigate effect of windowing on asynchronous breathing. Like previously, we only analyze asynchronous breaths, and then compare the estimated $C_{rs}$ of our algorithms without windowing and with rolling median windowing (Figure 10). Our findings show that windowing is able to reduce impact of asynchronous breathing by utilizing prior synchronous breaths to smooth estimated $C_{rs}$. Combined with results of (Figure 9), we see windowing is most effectively used on a per-breath basis during asynchronous breathing. This may mean that windowing is best used during real-time estimation of $C_{rs}$ in clinical practice.

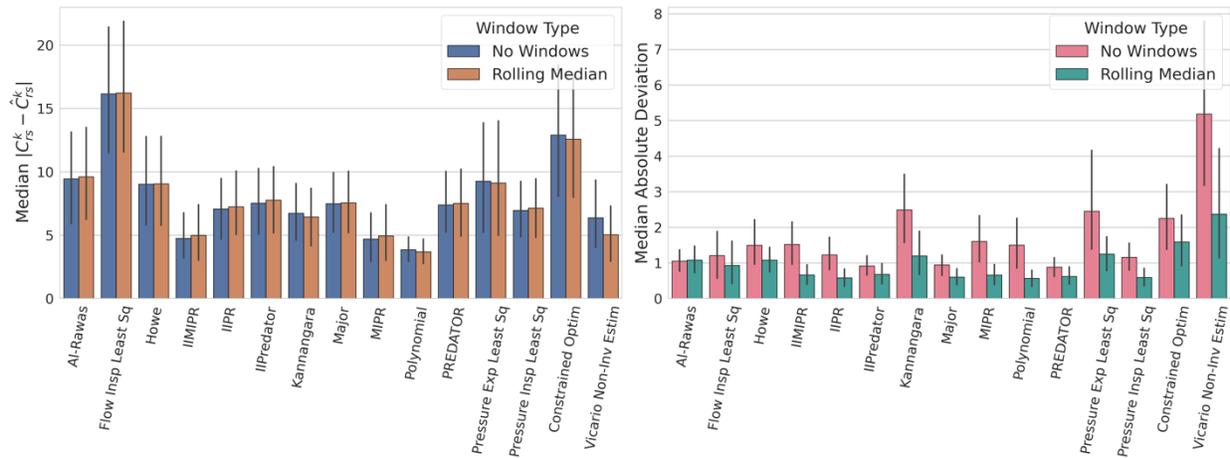

Figure 9: **A.** Shows the effect of windowing on absolute difference (AD) between etimated and true compliance. We were unable to find an algorithm that was statistically improved using windowed difference. **B.** Shows effect of windowing on median absolute deviation. Note: 95% CI is shown in barplot whiskers

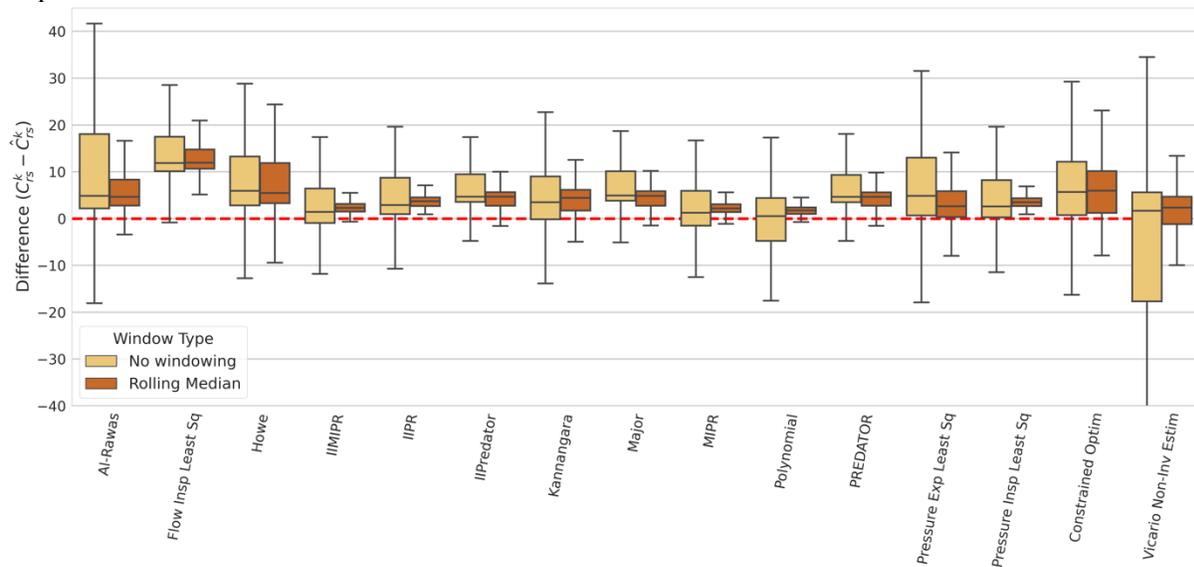

Figure 10: Shows the effect of no windowing (yellow) and rolling median windowing (orange) when examined at breath level. We first only choose asynchronous breaths across our entire dataset.

**Without Mode Specificity**

We also hypothesized there may be certain characteristics of mode-specific algorithms (Table 1) that translate to modes outside those they were developed in. So we investigate the utility of analyzing VM-specific algorithms in modes they were not tested for. In our first investigation, we compute the windowed, per-patient results for all PC/PRVC breaths with VC mode-specific algorithms, and compare them to baseline results derived our 2 PC/PRVC-specific methods (Table 1, Figure 11). Our findings show us that the polynomial method, MIPR, and IIMIPR outperform both PC/PRVC-specific algorithms in per-patient median AD, and the flow-targeted inspiratory least squares method dramatically underperforms pressure-targeted least squares. Our results imply that even though VC algorithms are clearly tailored for specific aspects of VC, that they possess some generic analytic properties that translate to PC/PRVC. Furthermore, the fact that flow-targeted inspiratory least squares dramatically underperforms pressure-targeted least squares is surprising, given that flow-targeted inspiratory least squares was mathematically formulated for PC/PRVC[7]. This shows that rearranging the single-chamber model to account for a specific VM can yield worse results than using the standard formulation.

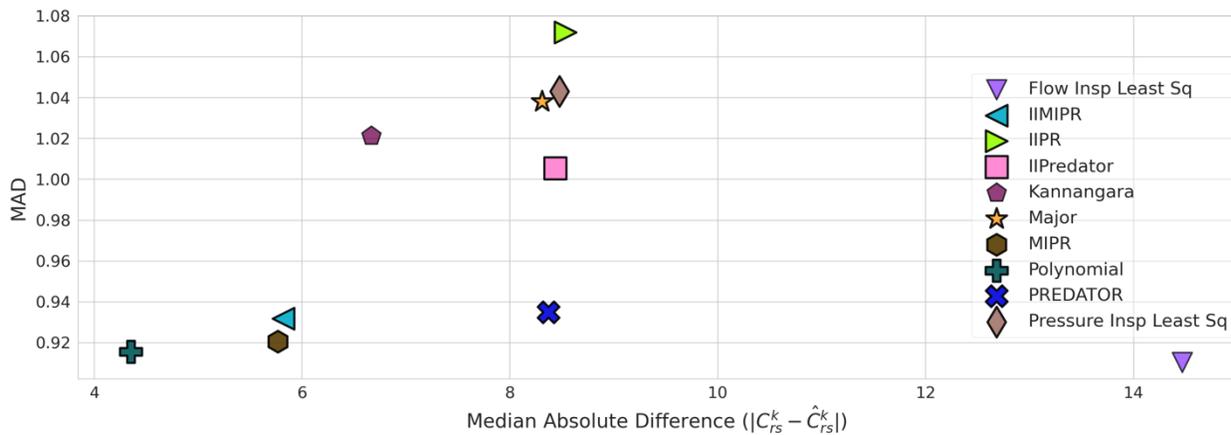

Figure 11: Shows the per-patient, windowed results of the VC specific algorithms applied to PC/PRVC. For comparison, we provide baseline calculations with PC/PRVC-specific algorithms.

We perform similar analysis on VC breaths. We analyze all VC breaths with our 2 PC/PRVC specific methods and provide baseline comparison using VC-specific algorithms. Unlike previously, none of the PC/PRVC specific methods outperform the best VC methods, but Kannangara's method is able to achieve relatively close performance to our best-performing VC algorithms. This experiment gives further evidence that there is generic utility to Kannangara's method even though its correction of flow will not function in VC-related asynchronies.

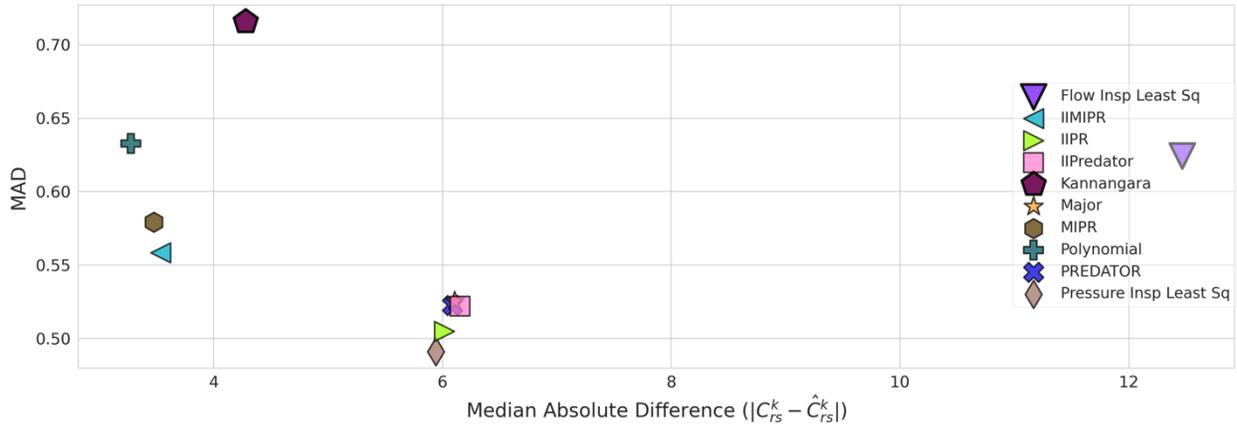

Figure 12: We find windowed, per-patient results for PC/PRVC algorithms when applied to VC. For comparison, we provide baseline calculations with VC-specific algorithms.

# DISCUSSION

In this paper, we have described our efforts to benchmark a comprehensive list of existing algorithms used to estimate $C_{rs}$ using the single-chamber model of a human lung. Our work showcases that the Polynomial method [34] works best in our patient population across all surveyed VMs (Figure 3, Figure 11). We also show that utilizing a median windowing strategy can improve overall consistency of algorithm results, and will function best when clinicians desire real time calculations for $C_{rs}$. Finally, we showed that all algorithms surveyed maintained good performance when applied to VMs outside those they were derived for. This result stands in constrast to existing assumptions that 1) mode-specific algorithms will only work in the VM they were developed for; and 2) the single chamber model needs to be rearranged to accommodate varying VMs [7]. In addition to these findings we have generated the largest available database of open ventilation data that can be used for benchmarking single-chamber algorithms. We hope that this contribution will speed up the development of future $C_{rs}$ estimators.

Our study provides the most comprehensive validation of available single-chamber algorithms along varying clinical scenarios, with the largest dataset of patients and breaths to date. Our study design differs from prior validation efforts by Redmond et al. where the authors analyze 6 algorithms with a dataset of 4 patients, where each patient receives sedation during monitoring[10]. Estimated $C_{rs}$ is then measured for 30 breaths before sedation, and 30 breaths after sedation, for a total of 240 breaths[10]. Our dataset was gathered over long periods of regular care during which time providers performed routine monitoring of $C_{rs}$ for a total of 39,660 breaths (see online supplement eTable 1). These differences in dataset size, study design, and the greater percentage of synchronous breathing in our dataset may explain some difference in results between ours and Redmond's study. For instance, when measuring $C_{rs}$, Redmond found that IIPREDATOR performed best when correcting for VC PVA. We note that IIPREDATOR was among the best performing algorithms in our study as well (Figure 6). However Redmond et al. did not evaluate expiratory pressure methods[5], [23] in their work. Our findings show using expiratory pressure methods[5], [23] function better than inspiratory pressure-targeted methods like IIPREDATOR when estimating $C_{rs}$ during moderate-severe FA. We also provide more granular discrimination into different types of PVA than Redmond[10]. Our work shows that all algorithms perform surprisingly well in periods of mild PVA such as during mild FA and DCA. It is only during more severe types of PVA that algorithms return variable $C_{rs}$ estimates. Redmond et al. did not perform this level of discrimination and treated all breaths in the period after sedation wore off as asynchronous[10]. Given the results of Redmond et al. and this study (Figure 6), it is likely that all subjects in Redmond et al. experienced moderate-severe FA after their sedation wore off[10].

Our results using VC algorithms in PC/PRVC, and vice-versa, requires elaboration. Algorithms that correct for PVA in VC like IIPR, MIPR, and the polynomial method will not function to correct PVA in PC/PRVC[21], [34], nor can Kannangara's method correct PVA in VC[7]. Our findings however, show there exist generic computational methods in these algorithms that allow them to perform well under VMs different than what they were designed for. For instance, Kannangara's implementation of their algorithm only utilizes a small portion of inspiratory data between two of the inspiratory "shoulders" to calculate $C_{rs}$[7]. Empirical evidence shows this implementation is significantly more effective than using the entire inspiratory portion of the breath that is done in flow-targeted inspiratory least squares[7]. Likewise, the way the polynomial method is calculating $C_{rs}$ may be yielding overall benefits that are mode-generic as well, however it is beyond the scope of this work to investigate the specific mechanism at play[34]. Furthermore, by comparing between results of pressure-targeted inspiratory least squares, and flow-targeted inspiratory least squares we find that rearranging the single-chamber model by VM does not yield improved $C_{rs}$ estimates (Figure 11). First, it is unclear if there is actual mathematic rationale for the rearrangement because VM does not change how lungs operate, and by proxy the single-chamber model, VM just changes how the ventilator delivers a breath. Second, rearranging the single-chamber model doesn't allow isolation of $R_{aw}$ and $C_{rs}$ components in the single-chamber model[7], [23]. This may contribute to the larger estimation errors seen in the flow-targeted inspiratory least squares method.

Although we have been able to provide further levels of clinical validation for the single-chamber model in different clinical scenarios our study also has limitations. First, our study is an observational, single-center study, and our IRB precluded any possibility of altering patient care. For a vast majority of patients we were also unable to capture valid plateau pressures above RASS of -3 when patients were truly spontaneously breathing. So, our results should only be viewed as applicable for patients with RASS$\leq -3$. We also experienced large confidence intervals in our estimation of windowing effects due to having a relatively small patient cohort of 18 subjects. Furthermore we are unable to make any claims about which $C_{rs}$ estimation algorithms work best under varying pathophysiologic states[5], [13]. Future work can add new open-source patient data to our published dataset so that the community can further improve confidence intervals for algorithm results, and understand performance of these algorithms under varying lung pathophysiology at different clinical centers.

# CONCLUSIONS

In this study we have provided further clinical validation of a set of 15 algorithms based on the single-chamber model of the lung. We show that under varying VMs, the polynomial method performs best at estimating per-patient respiratory compliance. We also explore how different algorithms perform under different types of PVA. We show that a simple windowed median can improve algorithm median absolute deviation and can serve as a useful smoothing function for real-time compliance estimation. Finally, we highlight the our benchmarking of algorithms in VMs other than what they were designed for, and open new questions for researchers to pursue in determining why certain algorithms perform well across VM. We hope our exploration will give scientists and clinicians more faith in performace of the single-chamber model under varying clinical scenarios and allow them to better tailor their compliance estimation techniques to patient physiology.

**Acknowledgments**

We are grateful for the support of everyone who helped at various stages of this project. We appreciate Dr. Brooks T Kuhn and Dr. Sarina Fazio for assisting with study design. Dr. Irene Cortes-Puch, Anna Liu, and Anna-Maria Nau for assisting with analysis. And finally, contributing respiratory therapists: Kenneth Ablog, Sarah Fonteno, Jonathan Lahola, Justin Griffiths, and Nazariy Siryy who assisted with the gathering of subject data and conducting end-inspiratory pause plateau checks..